# Classical entanglement

Douglas G. Danforth

danforth@csli.stanford.edu

Center for the Study of Language and Information, Stanford University, Stanford CA 94305

Classical systems can be entangled. Entanglement is defined by coincidence correlations. Quantum entanglement experiments can be mimicked by a mechanical system with a single conserved variable and 77.8% conditional efficiency. Experiments are replicated for four particle entanglement swapping and GHZ entanglement.

PACS numbers(s): 03.65.Ud, 03.65.Ta

## I. INTRODUCTION

The theory of matter and radiation developed from 1900 and called quantum mechanics has many features that are at variance with everyday experience. Erwin Schroedinger [1] coined the phrase 'entanglement' to describe the seeming nonlocal interdependency between the parts of a quantum system. It is generally believed that entanglement is "a resource which has no classical analogue" [2] and "The phenomenon of having correlated, or, in the language of quantum mechanics, entangled states separated by space, is one of the quintessential features in quantum mechanics and it has no analogue in classical physics" [3] . The reason for such belief stems from the work of J. S. Bell [4]. Bell's theorem is widely believed to apply to all classical systems, however, that it only applies to *lossless* systems. A system is lossless if every object impinging upon a detector causes a detection event, otherwise it is *lossy*. The current state of affairs is depicted in Table 1.

|  | quantum | classical |
|---|---|---|
| lossless | MAYBE[1] | NO[2] |
| lossy | YES[3] | YES[4] |

Table 1. When entanglement is possible

[1]  To date lossless (>82.8%) quantum experiments have not been achieved and hence the entry MAYBE  (there are claims [5] that the loophole is closed, however, there are doubts [6]).
[2]  Bell's results proves that no lossless classical system can reproduce the theoretical correlations of bi-partite quantum systems.
[3]  Lossy experimental results have indeed reproduced the predicted correlations of

        entangled quantum systems.
4. Lossy classical systems can also reproduce the predicted correlations of entangled quantum systems.

It is the purpose of this paper is to discuss lossy classical systems and show that such systems can mimic quantum correlations. The word 'mimic' is used to avoid the claim that the classical system to be introduced represents the physical processes underlying the quantum system. This paper shifts the debate from that of quantum *or* classical to that of quantum *and* classical systems. It puts them on an equal footing. Quantum systems are not unique in their ability to produce high space-like correlations.

The classical system introduced here exploits the detection efficiency loophole investigated by CHSH [7] and hence is not governed by Bell's inequality [4]. The mechanical system uses a single *conserved* variable to reproduce quantum bi-partite correlations. In this paper any classical system that reproduces quantum coincidence correlations of an entangled quantum system is called entangled.

In a recent article entitled "Experimental long-lived entanglement of two macroscopic objects" [8] it is said that $10^{12}$ atoms have been entangled. The current paper shows that classical macro states of very large numbers of atoms can be entangled.

This paper is organized as follows. In section II general classical principles are introduced which are sufficient to reproduce quantum correlations. Those principles are then specialized to a mechanical rod system. The rod system is then re-represented in cylindrical coordinates. In section III the theoretical and experimental conditions for quantum bi-partite measurements are described and the cylindrical model is shown to satisfy them. The cylinder model is then related to theoretical and experimental efficiency constraints. Section IV uses the cylinder model to duplicate experiments for several well known photon quantum cases. Section V concludes with a discussion of the significance of classical models.

## II. PRINCIPLES OF ENTANGLEMENT

It is possible to abstract three principles that lead to entanglement.

    Locality:        Interactions are local.
    Conservation:  Conservation laws apply.
    Detection:     State determines detection.

State dependent detection of systems asserts that only certain combinations of local object state and local detector state give rise to detection of the local object. For a given object state the detector state must be *commensurate* with it. With actual physical experiments (classical or quantum), a great deal of effort is needed to design and make a detector that is appropriate to the phenomenon under investigation. Detection does not come naturally. For example, detection of x-rays by inferred detectors is not expected. Scattering of optical photons by air does not happen and our eyes can not see the photons that are scattered. There are many such examples. They are usually considered unremarkable. Detectable states become remarkable when unexpected local degrees of freedom arise that affect detection.

There are consequences arising from these principles. *Coincidence* measurements are affected by loss. Bias is introduced by the rejection of good observations simply because another system detected nothing. This rejection of data is *nonlocal*. The outcome of a distant system determines the validity of local data. Statistics in ignorance of a distant system are different from data arising from coincidence. This bias need not be conspiratorial [9]. The anthropic principle [10] can be invoked to explain the way things are. The understanding that coincidence biases results is part of, but not sufficient to resolve, the question of quantum entanglement.

Conservation laws affect *correlations*. If a system splits spontaneously into two parts, the separate parts are correlated due to conservation. Separated classical systems are correlated by having been in local contact at some time in their past (or through intermediaries).

Locality, conservation laws, and state dependent detection are sufficient to reproduce quantum mechanical coincidence correlations using classical systems.

## A. Mechanical rod system

Consider a source that produces macroscopic rods of fixed length. Each rod is broken into two pieces and the pieces are hurled at separate units. The two pieces have the same constant orientation (polarization) which is perpendicular to the direction of motion.

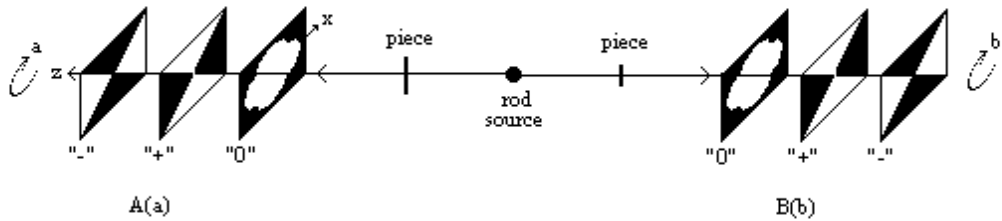

Figure 1. A mechanical model of rods, sheet metal filters, and detectors.

A filter is made of sheet metal cut in the form of a four leaf clover. If a piece has the right orientation and is not too long it will pass through the filter and hit one of two detectors ('+' or '-' ). If the piece is too long it hits the sheet metal (labeled '0') and is not detected. With the right choice of break point distribution and clover shape the quantum mechanical two particle correlation is reproduced (complete error analysis of this mechanical system would include: the error in non orthogonal orientation of a piece to its direction of motion; the possible non zero angular momentum of a piece; the non alignment of the direction of propagation to the center of a detection unit; the finite width of a rod to its length and the size of a cut in the '+' detector allowing passage through to the '-' detector. In this paper we assume that all of these effects can be made small relative to the overall response of the system.)

The three components of the mechanical unit can be considered a polarizing beam splitter (PBS) with two channels of detection. The mechanical PBS introduces loss into the system but only for rods not oriented along the orthogonal polarization directions. The PBS passes 81.8% of the pieces incident upon it. The detectors are 100% efficient. Note that this is in contrast to most discussions of *detector* efficiency. Here the loss is in the process of splitting the beam into two channels (forced choice).

The efficiency of an *optical* PBS is usually measured by examining the transmittance and reflectance of p and s polarized beams. For example:

    Transmittance
        p polarized beam:   >95%
        s polarized beam:   <0.01%
    Reflectance
        p polarized beam:   <5.0%
        s polarized beam:   >99.8%

Table 2. Specifications of an Oriel Instruments 26090 Polarizing Beam Splitter Cube.

If we make the assumption that rod pieces are polarized in only one direction (s and p beams are precise) then the corresponding specifications for the mechanical unit of Figure 1 are

    Transmittance(-)
        p(-) polarized beam: 100%
        s(+) polarized beam: 0%
    Reflectance(+)
        p(-) polarized beam: 0%
        s(+) polarized beam: 100%

Table 3. Specifications of mechanical PBS

There is *no loss* for the mechanical system for these special orientations. If the s and p beams are not precise but spread within $\pm 45°$ of the precise orientation, then Table 3 still holds for the mechanical model if the mechanism that produced the polarized pieces have an angular dependence on the piece lengths that falls within the radius of the sheet metal filter.

### B. Cylinder model

The response function, A, of a mechanical detection unit can be depicted on the surface of a cylinder where distance around the cylinder corresponds to piece orientation, $\lambda_1 = \theta$, and distance along the cylinder corresponds to half the piece length, $\lambda_2 = L/2$. The cylinder is scribed with a rectified sine wave above its mid line.

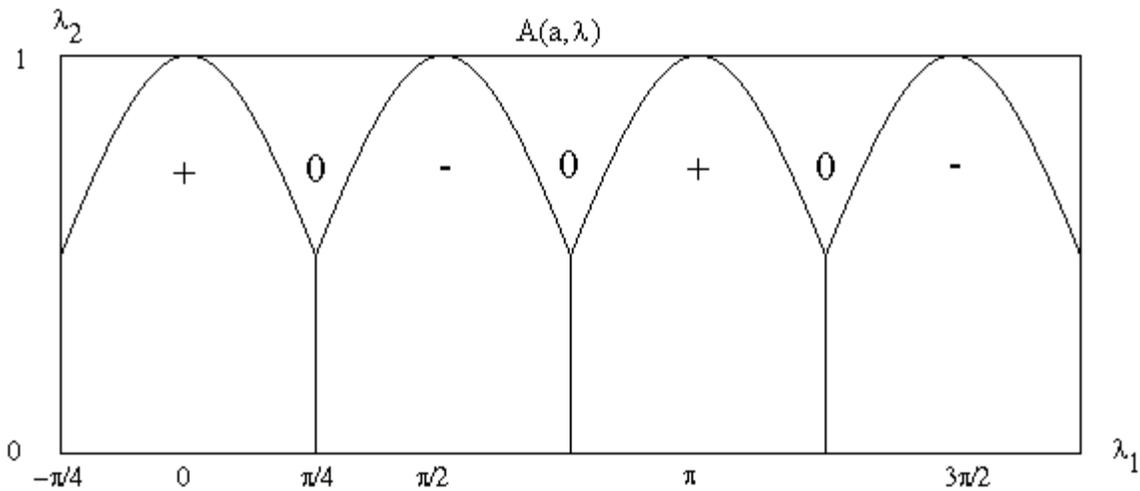

Figure 2. Cylinder Model (photons)

For photons there are four lobes. For electrons there are two. The overall rotation of the cylinder specifies the detector angle where the axis of the cylinder is collinear with the source. Both detectors are identical. In this cylindrical form, the distribution of variables is *uniform* over the cylinder. The region above each lobe is clear to signify no detection, '0'. The cloverleaf distribution is just a coordinate transformation of the uniform cylinder distribution where the base of the cylinder is deformed to a point and the scallops spread out and flattened onto the plane.

The cylinder model is a winner-take-all form of the Malus law for crossed optical polarizers. Each channel is governed by a cosine squared law coupled with a binary decision that chooses the channel with the greater probability. The periodicity of the two representations are related by the identity $\cos^2\theta = (1+\cos2\theta)/2$.

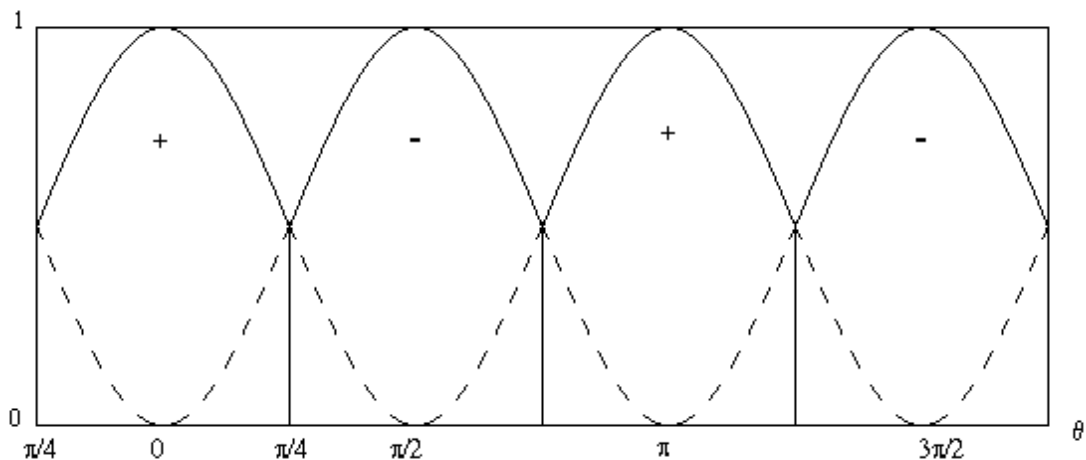

Figure 3. Orthogonal polarizers with winner-take-all Malus law.

For the cylinder model the sum of the length of the two pieces is conserved hence the joint behavior of the two detectors is best represented by conceptually *flipping* one detector (B) about its mid line ($\lambda_2=½$) with respect to the other (A) and overlaying them. Motion of a point in the $\lambda_2$ direction along the combined cylinders increases the value of the piece length for one particle and decreases it for the other. Hence a single point on the flipped and combined pattern represents both pieces.

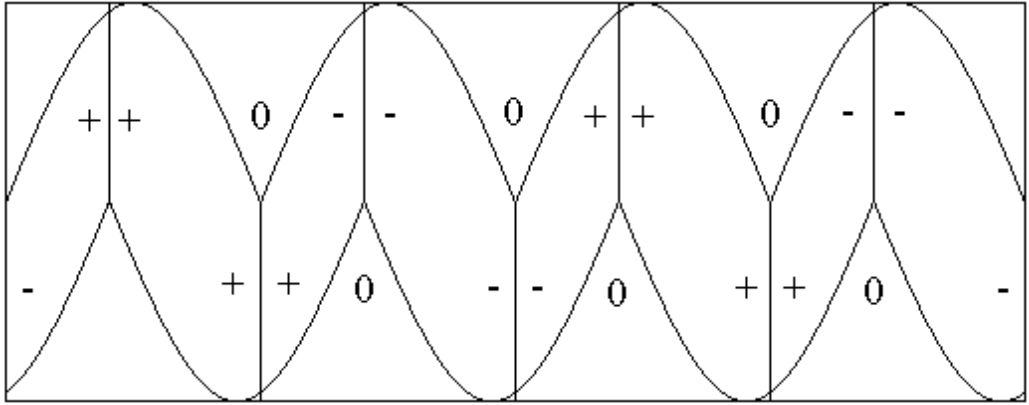

Figure 4. Joint detector response $A(a, \lambda)B(b,\lambda)$ for an arbitrary a and b orientation.

Points with signs from both cylinders are points of coincidence. The total area of coincidence is independent of the relative rotation angle, a-b, between the two cylinders. The integrated sum of signed values coincident over the surface of the cylinders reproduces the quantum mechanical 2 particle correlation as a function of relative cylinder rotation.

### III. MATHEMATICAL FORMULATION
#### A. Quantum bi-partite expectations

There are 9 conditions that must be satisfied by a classical system to reproduce quantum mechanical two particle correlations. These conditions are joint expectations of powers of the detector outcomes. The form $<X>$ is written as the expectation of a random variable X. As with J.S. Bell, the notation $A(a, \lambda)$, $B(b, \lambda)$ is used for the outcome of unit I and II respectively. Unlike Bell, the values that these variables can attain include 0, signifying no detection. The step in Bell's proof that relies on $B^2=1$ is not applicable for lossy systems. To see this clearly consider Figure 5 where the hidden variable space, $\Lambda$, has been divided into 9 regions depending upon the joint outcomes of $A(a, \lambda)$ and $B(b, \lambda)$. The left figure depicts a possible partitioning when the detector angle for A is a and for B is b. The right figure depicts another partitioning when a and b take new values a' and b'.

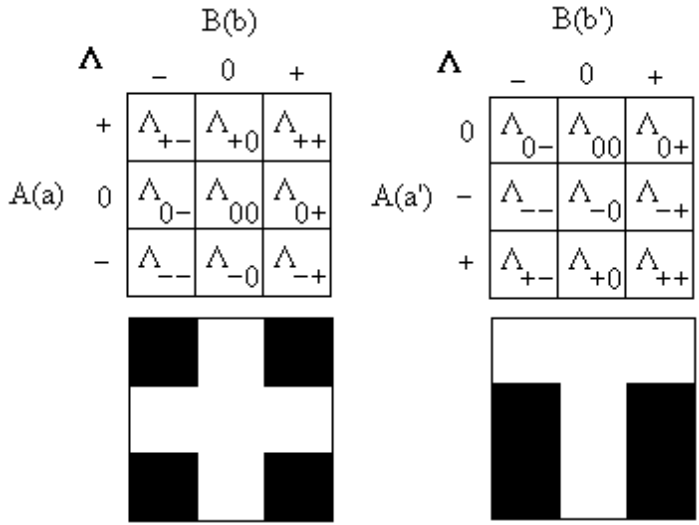

Figure 5. Partitioning of hidden variable space for (a,b) and (a',b') settings.

The all black regions are those regions for which a nonzero response occurs for *both* detectors. For lossless systems the whole $\Lambda$ space is black but for lossy systems the black regions, in general, move as a function of detector orientation. By conditionalizing on joint detection (restriction to black regions) the probability density $\rho(\lambda)$ used by Bell becomes a function of detector orientation (a, b), a situation which Bell precisely wished to avoid. Note that nothing strange or special is occurring for these lossy systems. The nonlocal dependency of the density function is a consequence of conditionalization on a detection system whose response is a function of the state of the local particle and the state of the local detector given that some particles are missed. All quantum detection experiments to date have a large degree of no detection (the section on efficiency below discusses this issue in more detail).

For lossy systems the concise specification of detector outcomes is given by

$$A^3 = A, \quad B^3 = B. \tag{1}$$

The 9 conditions are:

$$E^{\mu\nu}(a, b) = \langle A^\mu(a) B^\nu(b) \rangle = \begin{pmatrix} 1 & 0 & S \\ 0 & Dr(a,b) & 0 \\ S & 0 & D \end{pmatrix}, \quad u,v = 0,1,2. \tag{2}$$

The function r(a, b) is the theoretical and experimental quantum mechanical two particle correlation for particles produced in a singlet state:

$$r(a, b) = (-1)^n \cos n(a-b), \; n = 2s, \; s = \text{spin of particles} \tag{3}$$
$$r(a, b) = -\cos(a-b), \text{ electrons} \tag{3a}$$
$$r(a, b) = +\cos 2(a-b), \text{ photons} \tag{3b}$$

The expectation matrix, E, is a statement of the experimental results taking into account the imperfect nature of

the detection process. In more detail the elements of the matrix are:

$0 = \langle A(a) \rangle = \langle B(b) \rangle$, constant for all a, b (zero mean, equal counts in '+' and '-' channels)
$S = \langle A^2(a) \rangle = \langle B^2(b) \rangle$, constant for all a, b (*singles* probability)
$D = \langle A^2(a)B^2(b) \rangle$, constant for all a, b (*doubles* probability)
$0 = \langle AB^2 \rangle = \langle A^2B \rangle$, constant for all a, b (zero mean with distant particle detected) (4)

In the case of lossless detection D=S=1.

The matrix, E, uniquely determines a probability matrix, P, for detection events:

$$P^{\sigma\tau}(a,b) = \Pr(A(a)=\sigma, B(b)=\tau) = \begin{pmatrix} \frac{D(1+r)}{4} & \frac{(S-D)}{2} & \frac{D(1-r)}{4} \\ \frac{(S-D)}{2} & 1+D-2S & \frac{(S-D)}{2} \\ \frac{D(1-r)}{4} & \frac{(S-D)}{2} & \frac{D(1+r)}{4} \end{pmatrix}$$
(5)

$\sigma, \tau = -1, 0, +1$.

The quantities S and D are constrained by the probabilities ( $0 \leq P^{\sigma\tau} \leq 1$ ) so that

$0 \leq D \leq S \leq 1$, $2S - 1 \leq D$, (6)
or
$S \leq 1/2 + D/2$. (7)

### B. Analysis of the cylinder model

The cylinder model satisfies the mathematical expectation values of the previous section. The pairing of identical positive and negative panels with uniform probability distribution over the cylinder guarantees equal expected counts in both channels of a detector and hence a zero mean:

$0 = \langle A(a) \rangle = \langle B(b) \rangle$ (8)

This expectation is invariant under detector rotation since regions of positive and negative response rotate as rigid bodies and maintain their probability mass.

The rigid rotation also guarantees that total probability of a detection, whether positive or negative, is conserved under rotation so that expected singles counts are constant:

$S = \langle A^2(a) \rangle = \langle B^2(b) \rangle$ (9)

To convince oneself that double counts are invariant under rotation, note that because the detection variable (piece length) is conserved the bottom band of B acts as a cover for the top scalloped part of A (and vice versa). Every detectable point of A within the scallops is always within some detectable point of B's band no matter what angle the band is rotated relative to the scallops, as such intersection of the detection sets is constant.

$$D = \langle A^2(a)B^2(b)\rangle \quad \text{(constant)}. \tag{10}$$

To show that the cosine dependence of the correlation is generated by the cylinder model, first consider the detectors aligned with a-b = 0. Consider a single positive rectangle of B completely enclosing a single positive scallop of A. The intersection of these two sets is just the total area under the scallop, which is also the contribution to the probability D by this rectangle. When the cylinders are rotated relative to each other, part of the scallop moves out of the rectangle. As a positive amount moves out, an equal amount of negative scallop from the next panel moves into the rectangle.

Let $f(x)$, $0 \leq f(x) \leq 1/2$, $0 \leq x \leq 1$, $0 = f(0) = f(1)$, be the bounding curve of a scallop. Let $F(x)$ be the area under the curve in the interval $[0, x]$. The net contribution to the correlation by the top half of a rectangle is

$$(-1)^n [(F(1) - F(x)) - F(x)] = (-1)^n [F(1) - 2F(x)].$$

The factor $(-1)^n$ specifies the negative overlap of panels for electrons when detectors are aligned and the positive overlap for photons. The total covariance is 4n times this quantity for the 4n half rectangles (top and bottom) in the overlap of the two cylinders.

The coincidence correlation, Q, (as used experimentally) is the expected product of detector outcomes normalized by the expected coincidence firing rate (which is constant and equal to D):

$$Q(x) = \langle A(a)B(b)\rangle / \langle A^2(a)B^2(b)\rangle, \quad \pi x = n(a-b), \tag{11a}$$

which for a general cylinder model with an arbitrary (piecewise continuous) form of scallop becomes:

$$Q(x) = (-1)^n [1 - 2 F(x)/F(1)], \tag{11b}$$

(factors of 4n and the constant uniform probability density of the hidden variables cancel).

One is now free to choose a functional form for F that yields the desired quantum mechanical correlation. Setting $Q(x) = r(x) = r(a,b)$. Solving for F gives

$$F(x) = \tfrac{1}{2}(1-(-1)^n r(x))F(1).$$

Hence

$$f(x) = F'(x) = -\tfrac{1}{2}(-1)^n r'(x) F(1) = \tfrac{1}{2}(-1)^n (-1)^n \pi \sin \pi x \, F(1)$$

or

$$f(x) = \tfrac{1}{2}\sin \pi x. \tag{12}$$

The constant $F(1)$ is equal to $1/\pi$.

## C. Efficiency of cylinder model

One can define three measures of efficiency: single, double, and conditional. Only the conditional efficiency can be measured unambiguously. The efficiency of a single detector, S, can not be determined without auxiliary assumptions since the absolute rate of particle impingement is not known. The same is true for the doubles count, D. The conditional efficiency (probability of detecting two particles given that one particle is detected) is independent of the absolute impingement rate (assuming linear detector response as a function of flux intensity):

$$C = D/S, \text{ conditional efficiency.} \tag{13}$$

C will be equal to unity if for every observation at detector I there is also an observation at detector II and vice versa. This corresponds to the line labeled "D = S" in Figure 6. Bell [4] considered the case D=S=1.

The surface area of a unit height cylinder in $\pi x = n(a-b)$ coordinates is 2n. When cut into 2n panels each panel has area 1. The doubles efficiency, D, is 2F(1) out of 1 or

$$D = 2F(1) = 2/\pi = 0.636 \tag{14}$$

The singles efficiency is:

$$S = 1/2 + F(1) = 1/2 + 1/\pi = 0.818 \tag{15}$$

The conditional efficiency is then:

$$C = D/S = 2F(1)/(1/2 + F(1)) = 4/(\pi+2) = 0.778. \tag{16}$$

This corresponds to the point labeled "Cylinder" in Figure 6. The line labeled "Clauser" corresponds to his upper bound on the conditional correlation of 0.828 [7]. It should be noted that Aspect's 1982 experiment was less than 1% efficient and is depicted by the line labeled "Aspect."

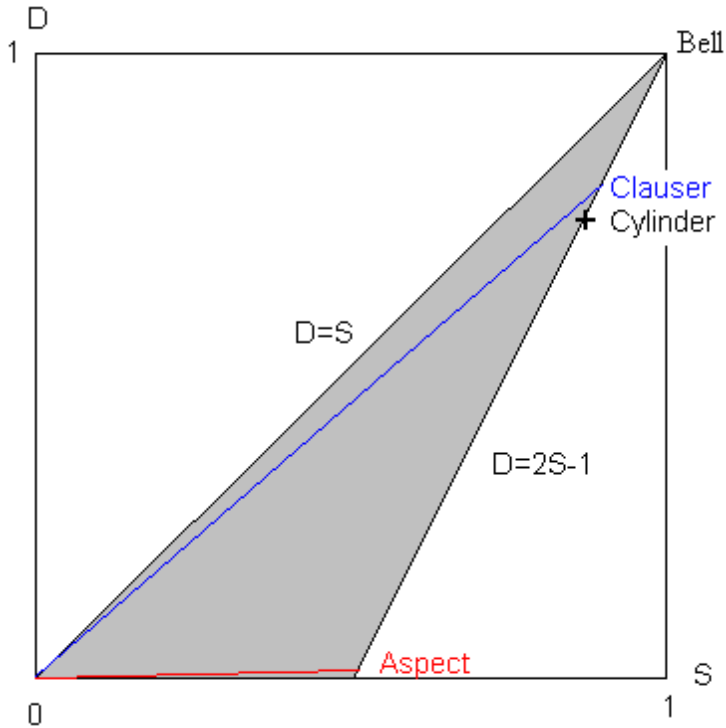
Figure 6. Detector efficiency constraints

Garg [11] and Fine [12] have raised the possibility that some experimental arrangement with N measurements angles for detector I and N measurement angles for detector II (N x N experiment) could refute local realism with some low detector efficiency. The cylinder model is valid for all continuous angles and shows that conditional detector efficiency for N x N experiments must be at least 77.8%. This differs by 5% from Clauser's 2x2 requirement of 82.8% and so effectively destroys the hope of a low efficiency experimental refutation of local realism.

## IV. REPRODUCTION OF EXPERIMENTS

Once a classical model has been found that mimics quantum 2 particle correlations one might assume that other multi particle correlations can be derived from it. This is the case for at least two widely discussed experiments.

### A. Two photon entanglement

By construction the cylinder model reproduces the bi-partite results of Freedman[13], Aspect [14], and others. Note that for parametric down conversion photons are selected to have orthogonal polarization. The rod model can accommodate this by simply constructing a source that emits pieces with orthogonal orientations.

### B. Four photon entanglement swapping

The cylinder model can reproduce the four photon entanglement swapping results of PBWZ's [15]. Their figure 2 is reproduced here for reference.

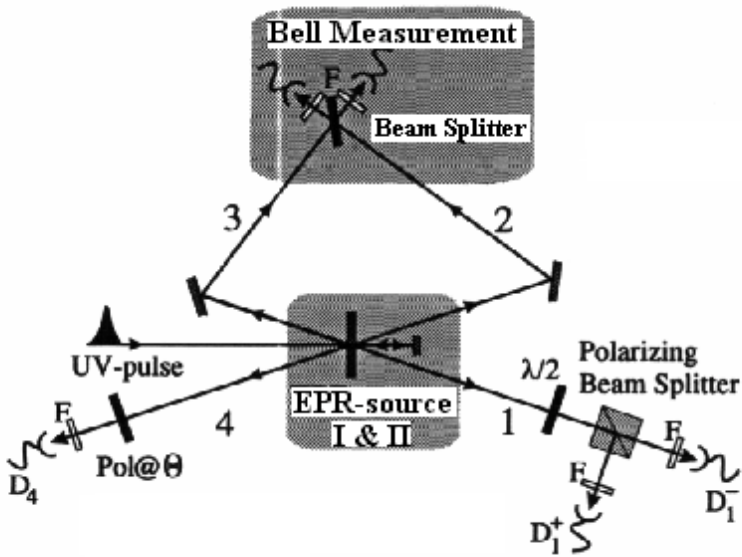

FIG. 2. Experimental setup. A UV pulse passing through a nonlinear crystal creates pair 1-2 of entangled photons. Photon 2 is directed to the beam splitter. After reflection, during its second passage through the crystal the UV pulse creates a second pair 3-4 of entangled photons. Photon 3 will also be directed to the beam splitter. When photons 2 and 3 yield a coincidence click at the two detectors behind the beam splitter, they are projected into the $|\Psi^-\rangle_{23}$ state. As a consequence of this Bell-state measurement the two remaining photons 1 and 4 will also be projected into an entangled state. To analyze their entanglement we look at coincidences between detectors $D_1^+$ and $D_4$, and between detectors $D_1^-$ and $D_4$, for different polarization angles $\Theta$. By rotating the $\lambda/2$ plate in front of the two-channel polarizer we can analyze photon 1 in any linear polarization basis. Note that, since the detection of coincidences between detectors $D_1^+$ and $D_4$, and $D_1^-$ and $D_4$ are conditioned on the detection of the $\Psi^-$ state, we are looking at fourfold coincidences. Narrow bandwidth filters (F) are positioned in front of each detector.

and their Fig 3 experimental results are reproduced below:

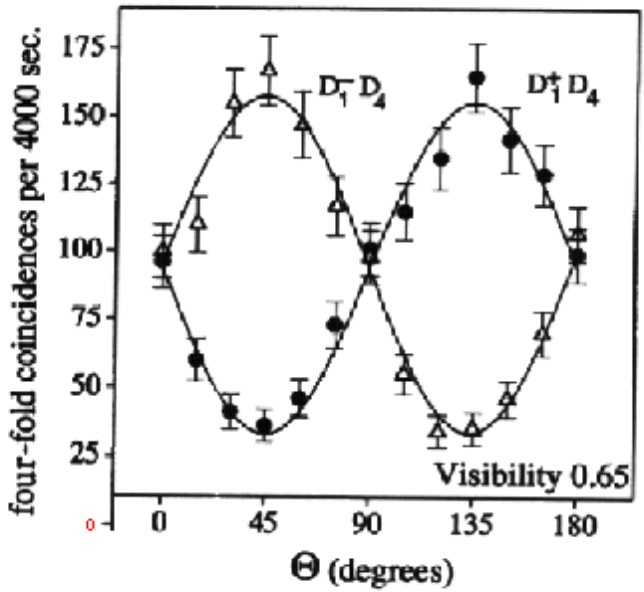

FIG. 3. Entanglement verification. Fourfold coincidences, resulting from twofold coincidence D1$^+$D4 and D1$^-$D4 conditioned on the twofold coincidences of the Bell-state measurement, when varying the polarizer angle Θ. The two complementary sine curves with a visibility of 0.65 ± 0.02 demonstrate that photons 1 and 4 are polarization entangled.

To assess the statistics of their experiment, the setup was simulated using the cylinder model. A sample run is shown in Figure 7. At each angle θ of detector 4 (taken as one channel of a mechanical PBS), 1800 groups of four photons were created. This number was chosen to approximate PBWZ's overall counting rate. Particles 1 and 3 of a group were generated with uniform random orientation and uniform random length. Particles 2 and 4 were made orthogonal to 1 and 3 (respectively) and their lengths made complimentary to their partners. Four-way coincidence counts were collected. The mean and standard deviations of these counts over 64 repetitions at each angle are depicted in Figure 7. The dotted lines are a least squares fit to a sine curve. For this setup the classical cylinder model has a visibility of 70.7% (√2/2) as compared to 65% for PBWZ.

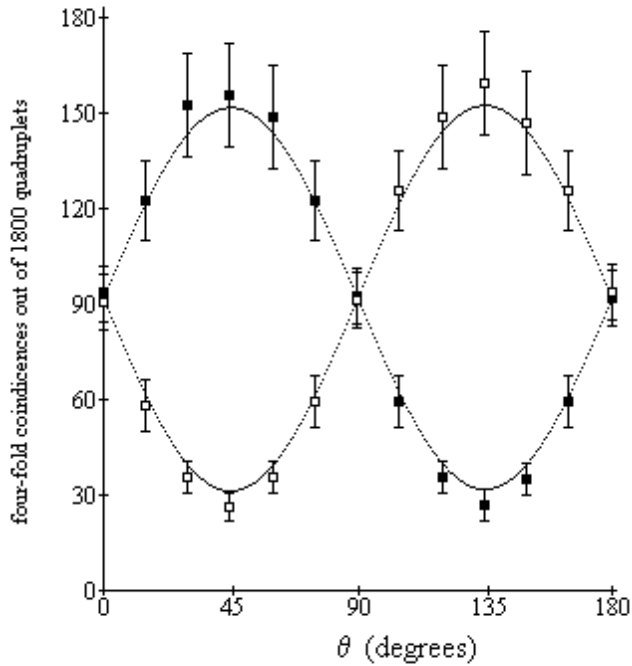

Figure 7. PBWZ simulation. Black points are D1⁻D4 counts and white points are D1⁺D4 counts.

Using the arguments of PBWZ one is drawn to conclude the mechanical system is performing entanglement swapping.

### C. GHZ entanglement

In a recent paper PDGWZ [16] discuss highly pure four-photon GHZ entanglement for the purpose of verifying the GHZ predictions and to demonstrate teleportation of entanglement with very high purity. The PDGWZ experimental setup is reproduced below

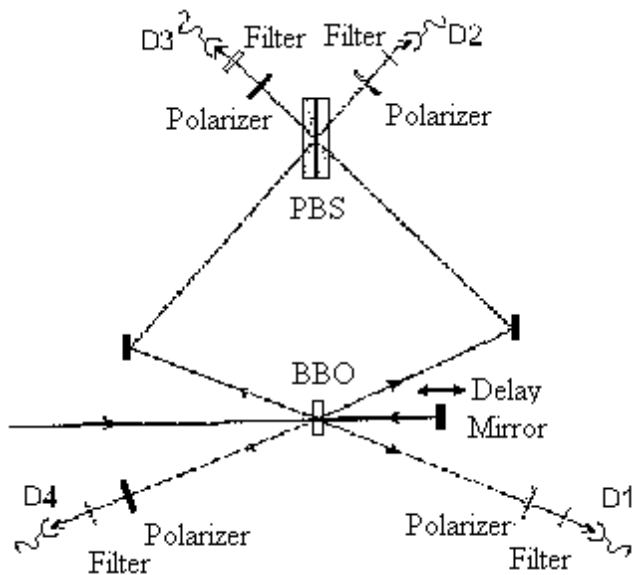

FIG. 2. Schematic of the experimental setup for the measurement of four-photon GHZ correlations. A pulse of UV-light passes a BBO crystal twice to produce two entangled photon pairs. Coincidences between all four detectors 1-4 exhibit GHZ entanglement.

Coincidence between D2 and D3 occurs when photons 2 and 3 are both V (vertically polarized) or both H (horizontally polarized). The PBS passes horizontal photons and reflects vertical ones. A VH or HV combination does not produce a coincidence count.

To reproduce this PDGWZ experiment, the PBS is duplicated in the mechanical setup (Figure 8) and data is recorded only when there is coincidence between (D$_2$', D$_3$') or (D$_2$", D$_3$").

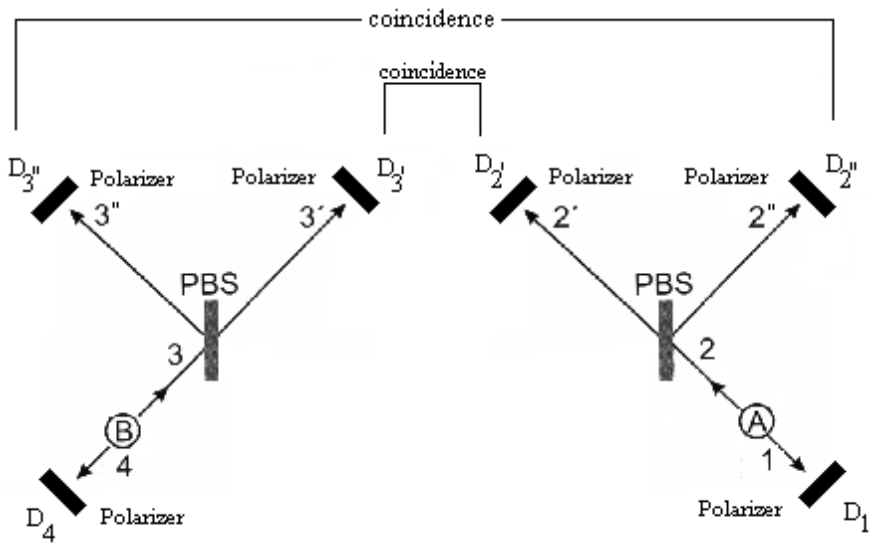

Figure 8. Mechanical GHZ experimental setup

Rods are produced at A and B (at the same time), are broken, and the pieces (1, 2, 3, 4) thrown. The piece orientation is uniformly random with complimentary pieces orthogonal to each other. Distribution of a piece length is uniform.

PDGWZ varies their 4 polarizers P1..P4 over all 16 binary combinations of horizontal or vertical orientation to test that their entangled final state |Ψ$^f$> has been obtained. In the classical mechanical setup polarizers P3" and P2' are forced to have the same orientation so as to correspond to the single polarizer P3. Similarly P3' and P2" have identical polarizations equal to P2. PDGWZ find that only HVVH and VHHV orientations produce coincidence counts and no others. Let us trace the pieces of the classical situation and show that this is also the case.

Let us initially set the polarizers to HVVH (1..4) for discussion. To be detected piece 1 must be within 45 degrees of H and hence piece 2 will be within 45 degrees of V since they are emitted at 90 degrees to each other. If piece 2 is not too long it will reflect from the PBS and become piece 2" (vertical pieces are reflected). Since 2" is within 45 degrees of V and P2"=P2=V (by HVVH setup) and 2" will be detected. Note that if P2" were to equal H it would not be detected and hence only P1=H and P2"=V or P1=V and P2'=H will produce coincidences. The same argument follows for P4 and P3', P3". Hence only HVHV, HVVH, VHVH, or VHHV detection can occur. But the added coincidence constraint that (D2', D3') or (D2", D3") that only *HH* and *VV* are detected. This leaves only HVVH and VHHV to occur in the classical experiment just as with the PDGWZ situation.

PDGWZ next demonstrate that their final state is in a coherence superposition by orienting the polarizers in two separate configurations (+45, +45, +45, +45) and (+45, +45, +45, -45) (in degrees). PDGWZ use a right hand rule [17] along the direction of each photon's propagation to specify positive angles. In these two different configuration (with zero time delay between photons 2 and 3 arriving at the PBS) they obtain two different four-fold coincidence counting rates. The visibility for these two rates is stated as being 0.79±0.06

The corresponding classical experiment goes as follows. Consider Figure 9 that depicts all possible discrete polarizations as 8 sectors labeled 1..8. Sectors with horizontal (vertical) polarization are labeled H (V). Sectors with +45 (-45) degree polarization are labeled "+" ("-"). If a particle has polarization H, it can fall into either a "+" or "-" sector and similarly for V polarization.

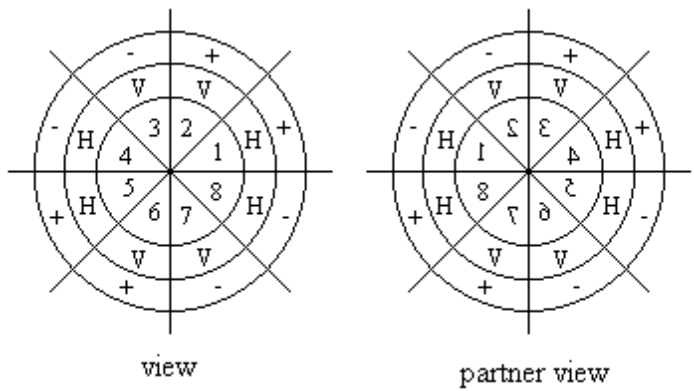

Figure 9. Polarization sectors for PDGWZ experiment

Pieces 2 and 3 are emitted and coincidence between (2', 3') or (2", and 3") is enforced. For configuration (+45, +45, +45, +45) each piece (2, 3) has a 50% chance of passing through their polarizers and being detected, (H1, H5) or (V2, V6).

Particles belonging to a pair have orthogonal polarization. 1H+ corresponds to 3V- if the particles are emitted in the same direction, however, they are emitted in opposite directions and so viewing 3V- from the partner's right hand coordinate system it becomes 3V+ (partner view). Table 4 shows the pattern of detections for all sectors (5..8 are the same as 1..4).

| particle sector | +detected | partner sector | +detected | both detected |
|---|---|---|---|---|
| 1(+) | yes | 3(+) | yes | yes |
| 2(+) | yes | 4(+) | yes | yes |
| 3(-) | no | 5(-) | no | no |
| 4(-) | no | 6(-) | no | no |

Table 4. Detection pattern for (+45, +45) polarizer pair

From this it is seen that if a particle is detected then so is its partner and there will be a net positive coincidence count for this configuration and so a net 4-fold coincidence count.

For the (+45, +45, +45, -45) it is necessary only to consider particle 3 with +45 and its partner 4 with -45 degree polarizer orientation. Table 5 shows the pattern of response.

| particle sector | +detected | partner sector | -detected | both detected |
|---|---|---|---|---|
| 1(+) | yes | 3(+) | no | no |
| 2(+) | yes | 4(+) | no | no |
| 3(-) | no | 5(-) | yes | no |
| 4(-) | no | 6(-) | yes | no |

Table 5. Detection pattern for (+45, -45) polarizer pair

From this it is seen that if a particle is detected then its partner is not detected and hence there will be a zero 4-fold coincidence. The GHZ result for this classical system therefore yields a visibility = (max-min)/(max+min) = 1.0 and so can be called strongly entangled.

Using PDGWZ's arguments it can be said the classical system is in a coherent superposition of HVVH and VHHV. It is interesting to note that the classical analysis of this PDGWZ experiment did not need the precise form of the cylinder model but only the disjointness of is response regions. If one considers PDGWZ as an adequate representation of GHZ entanglement then lossy classical systems perform GHZ entanglement.

## V. DISCUSSION

The thrust of this paper is that lossy classical systems are *equivalent* to quantum systems *relative* to coincidence correlations. Both systems have the same power. The systems may not be identical, but they are equivalent. Even if future highly efficient quantum experiments support the theoretical correlations the results of this paper still hold. It will still be possible to simulate the experimental results with a lossy classical system.

**Teleportation and communication**
If it is assumed that the PBWZ and PDGWZ experiments characterize entanglement swapping and quantum teleportation then it has been shown that lossy classical systems can also perform entanglement swapping and teleportation with efficiency higher than that reported in the current photon experiments. This has consequences for the interpretation of quantum communication.

There is no *collapse of a wave function* when a measurement is made on a rod by either A(lice) who is in communication with B(ob) or by Bob. The order of measurement or rapid change in the detector orientation do not matter since results are determined at the time of impingement. Even so, the rod pieces are entangled as evidenced by their correlation.

**State**
A rod's orientation takes on a continuum of values. The outcomes '+' or '-' are assigned by the measurement process. These discrete outcomes say very little about the orientation of the rod. They are a meager representation of the actual state of the rod. They are a projection of its state by the measurement device.

The wave function ψ of a quantum system is frequently called its *state* [19], [20], [21]. It is actually the *probability amplitude* of measurement outcomes. The rod model emphasizes the fact that such a probabilistic specification can not represent the classical state of a specific piece but only a distribution over an ensemble of pieces (it is suggested that the phase *amplitude of the system* be used referring to ψ).

**Quantum computing**

Peter Schor [18] has discussed the computational capabilities of quantum systems. Because of his work it is believed that classical systems are less powerful (<) than quantum systems. The current paper states that lossy classical systems are equivalent (~) to quantum system,

    classical          <     quantum     (Shor)
    lossy classical   ~     quantum     (Danforth),

hence it seems that

    classical          <     lossy classical,

which is an obvious contradiction.

Either the equivalence relationship based on coincidence correlation (entanglement) is irrelevant to computational power or it is false that classical systems are less powerful than quantum systems.

**EPR**

It is possible to reinvestigate the issues raised by Einstein, Podolsky, and Rosen (EPR) [22]. The assumption of predicting "with certainty" the outcome of a distant measurement "without disturbing the system" can not be realized for lossy systems. A detection at one does not predict a detection at the other. Only if both detectors fire can one say "with certainty" what the outcome of the distant detector will be when the detectors are aligned. But this knowledge is "after the fact". It is ironic that for this blatantly classical system it is not possible to apply the EPR arguments even though the classical system reproduces quantum entanglement. Hence Einstein's goal of showing the incompleteness of quantum mechanics can not be reasoned from EPR for lossy systems.

## ACKNOWLEDGEMENTS

I wish to thank Bill Taylor in the department of Mathematics at Canterbury University, New Zealand for posting to the internet (in 1992) his clear analysis of Mermin's [9] comments. His analysis crystallized thoughts begun during seminars at Stanford University on the foundations of physics conducted by Professor Patrick Suppes in the '70s.

I also wish to thank John F. Clauser for a pleasant afternoon in the summer of 1992 at Berkeley discussing this work and his encouragement of its publication.

---